\documentstyle[12pt]{article}

\def\beq{\begin{equation}}
\def\eeq{\end{equation}}
\def\lesim{{\buildrel <\over\sim}}
\def\gesim{{\buildrel >\over\sim}}

\begin{document}
\pagestyle{empty}
$\ $
\vskip 2 truecm

\centerline{\bf BLACK HOLES FROM NUCLEATING STRINGS}
\vskip 1 truecm
\centerline{Jaume Garriga$^{1,2}$ and Alexander Vilenkin$^{1,3}$}
\vskip .5 truecm
\centerline{\em $^1$Tufts Institute of Cosmology,}
\centerline{\em Department of Physics and Astronomy,
              Tufts University, Medford, MA 02155}
\vskip .3 truecm
\centerline{\em $^2$Institute for Theoretical Physics,}
\centerline{\em University of California}
\centerline{\em Santa Barbara, California, 93106-4030}
\vskip .3 truecm
\centerline{\em $^3$Lyman Laboratory of Physics, Harvard University,}
\centerline{\em Cambridge, MA 02138}
\vskip 2.5 truecm

\begin{abstract}

We evaluate the probability that a loop of string that has
spontaneously nucleated during
inflation will form a black hole upon collapse, after the end of inflation.
We then use the observational bounds on the density of primordial black
holes to put constraints on the parameters of the model. Other constraints from
the distortions of the microwave background and emission of gravitational
radiation by the loops are considered. Also, observational constraints on
domain wall nucleation and monopole pair production during inflation are
briefly discussed.

\end{abstract}

\clearpage
\pagestyle{plain}
\section{Introduction}
\label{introduction}

It has been proposed \cite{baal91,bavi92} that loops of string
can spontaneously nucleate during an inflationary period of expansion in
the early universe. This is a quantum tunneling process,
somewhat analogous to the spontaneous nuclealion of spherical bubbles of
true vacuum in the problem of false vacuum decay \cite{co85}.
It has been shown that loops can nucleate at considerable
rates provided that their tension $\mu$ is not much larger than $H^2$
(here $H$ is the expansion rate during inflation) and,
in this case, the distribution of loops emerging from inflation may have
significant cosmological consequences.

In this scenario, loops nucleate with sizes of the order of $H^{-1}$,
and are subsequently stretched to large sizes by the
inflationary expansion. After inflation, in the radiation dominated era,
the loops eventually fall
within the Hubble radius and start oscillating under their
tension.

Because the instanton describing the nucleation of a loop is maximally
symmetric, the nucleated loops tend to be nearly circular. It is well known
that an exactly circular loop would form a black hole upon collapse
\cite{vi81,ha89}. However, the nucleated loops will not be {\em exactly}
circular. The reason is that, during inflation,
quantum fluctuations on the string on
subhorizon scales generate shape instabilities as their wavelengths become
larger than the horizon \cite{gavi91,gavi92} (see also \cite{tu88}).
The question then arises of how many of the
nucleated loops will be circular enough to form a black hole after the end
of inflation. This is the problem we address in the present paper.

The plan of the paper is the following.
In Section 2 we calculate the probability of black hole formation from a
nucleating string.
In Section 3 we use this result and the
observational bounds on the abundance of primordial black holes to put
constraints on the parameters of the model (which are essentially $\mu$
and $H$.) Additional constraints from the distortions in the microwave
background and the emission of gravitational waves by the strings are
considered.
In Section 4, we briefly discuss observational constraints on
domain wall nucleation and monopole-antimonopole pair production,
which can also occur
during inflation \cite{baal91}.
We summarize our conclusions in
Section 5. Some details are left to the appendices.
Appendix A deals with linearized perturbations on a circular loop of string
in flat space, and in Appendix B we consider the effect of damping
processes on the probability of black hole formation.

\section{Probability of black hole formation}
\label{formation}

In this section we estimate the probability that a loop of string that
has nucleated during inflation will form a black hole once it collapses under
its tension, after the end of inflation.

The classical worldsheet of a circular loop of string that has nucleated during
inflation is given by \cite{baal91}
\beq
R(t)=H^{-1}\sqrt{e^{2H(t-t_0)}+1},\label{wsheet}
\eeq
where $R$ is the physical radius of the loop, $H$ is the expansion rate of the
inflationary universe and $t$ is the usual cosmological time in the flat
Friedman-Robertson-Walker coordinates. The parameter $t_0$ can be
interpreted (in a loose sense) as the time at which the loop nucleates.
Throughout this paper, we shall work in the approximation in which the
strings are infinitely thin, although nucleation of strings whose thickness
is not much smaller than $H^{-1}$ is also possible \cite{bavi92}.

The loops nucleate with a size of the order of the horizon size
$R\sim H^{-1}$, and afterwards they are stretched by the inflationary
expansion. For $R>>H^{-1}$ they grow proportionally to the scale
factor
$$
R(t)\approx H^{-1}e^{H(t-t_0)}.
$$
After inflation, in the radiation dominated era, the loops continue to be
stretched by the expansion of the universe, $R(t)\propto t^{1/2}$, until
they enter the cosmological horizon, at  $t\sim R$. Soon after the loop comes
within
the horizon, the effects of expansion become negligible and the loop starts
behaving approximately as it would behave in flat space.

Let $R_c$ be the radius of a particular loop at horizon crossing. The mass of
this loop is $2\pi R_c\mu$, where $\mu$ is the string tension. The
Schwarzschild radius corresponding to this mass is
\beq
R_s=4\pi G\mu R_c.
\eeq
As the loop shrinks under its tension, its rest mass is
converted into kinetic energy, so that the total energy of the loop remains
constant. It is clear that if the loops were exactly circular, they
would all eventually shrink to a size smaller than $R_s$, thus forming black
holes (of course this argument assumes that $R_s$ is larger than the
thickness of the string,
which will always be true for sufficiently large loops.)

However, loops nucleated during inflation will not be exactly circular.
Quantum fluctuations around the symmetric solution (\ref{wsheet})
will cause small departures from the circular shape, and this
will determine the probability of black hole formation.
Consider a circular loop of radius $R(t)$ lying in the $z=0$ plane and
centered at the origin. Using cylindrical coordinates $(\rho,\theta,z)$,
we can parametrize a perturbed (non-circular) loop as follows:
\beq
\rho(\theta,t)=R(t)+\Delta_r(\theta,t),\label{parametrize}
\eeq
$$
z(\theta,t)=\Delta_t(\theta,t).
$$
That is, we decompose the perturbations into a radial part $\Delta_r$
parallel to the plane of the unperturbed loop, and a transverse part
$\Delta_t$ perpendicular to that plane. In eq.(\ref{parametrize}),
$\rho$ and $z$ are physical coordinates (rather than co-moving.)

To study the evolution of a loop, we shall divide its history into
three epochs:
\begin{description}
\item $a$- The loop nucleates and then expands during inflation.
\item $b$- After inflation, while the loop is larger than the cosmological
horizon it is stretched by the expansion until it crosses the horizon.
\item $c$- Once the loop comes within the horizon, it shrinks under
its tension.
\end{description}
We will follow the evolution of the perturbations
$\Delta_r,\Delta_t$ through the epochs $a$, $b$ and $c$. For this it will
be useful to expand them in Fourier modes:
\beq
\Delta_r=\sum_{L=2}^{\infty}\left[
\Delta_{r,L}^{(1)} {\cos L\theta\over\sqrt\pi}+
\Delta_{r,L}^{(2)} {\sin L\theta\over\sqrt\pi}\right], \label{fourier}
\eeq
and similarly for $\Delta_t$. Note that the sum does not include the modes
with $L=0$ and $L=1$. This is because, as it was shown in Ref.\cite{gavi92},
such modes do not correspond to true perturbations, but only to spatial
rotations and spacetime translations of the unperturbed solution
(\ref{wsheet}).

{\em a- Evolution during inflation.} In the process of nucleation,
$\Delta^{(i)}_{\lambda,L}$ ($i=1,2$; $\lambda=t,r$) have to be treated as
quantum variables. The linearized theory of quantum fluctuations around the
classical solution (\ref{wsheet}) was studied in Ref. \cite{gavi92}, using a
covariant formalism. The worldsheet
(\ref{wsheet}) has the internal geometry of a 1+1 dimensional de Sitter
space \cite{baal91}, and it was shown that the
perturbations $\Delta_r$ and $\Delta_t$ behave as two uncoupled scalar
fields of tachyonic mass $m^2=-2H^2$ `living' in this lower dimensional de
Sitter space. The symmetries of the problem suggest that when the string
nucleates, the fields $\Delta_r,\Delta_t$ should be in a de Sitter invariant
quantum state (see also Ref.\cite{vavi91}). This state was constructed in
Ref.\cite{gavi92} using the Heisenberg picture. The corresponding Schrodinger
picture wave functional
$\Psi(\{\Delta^{(i)}_{\lambda,L}\})$
can be obtained using well known manipulations  (see e.g. \cite{vavi91}).
Since we are dealing with a linearized theory, the probability distribution
associated with $\Psi$ has the Gaussian form
\beq
{\cal P}(\{\Delta^{(i)}_{\lambda,L}\})=|\Psi|^2=
\prod_{\lambda,i,L}(2\pi\ \sigma_{\lambda,L}^2)^{-1/2}
\exp\left[{-(\Delta^{(i)}_{\lambda,L})^2\over 2\ \sigma_{\lambda,L}^2}\right],
\label{probability}
\eeq
where the standard deviations $\sigma_{\lambda,L}$ are given by \cite{gavi92}
\beq
\sigma_{\lambda,L}^2(t)\equiv<\Psi|(\Delta^{(i)}_{\lambda.L})^2|\Psi>=
{1\over 2\mu L}\left[{H^2R^2(t)\over L^2-1}+1\right].\label{sigma}
\eeq
For wavelengths much larger than the horizon, $HR>>1$, we have
$\sigma_L\propto R$, i.e., the amplitude of the perturbations is conformally
stretched by the expansion (recall that $R$ grows like the scale factor
during inflation). For wavelengths much smaller than the horizon the second
term in (\ref{sigma}) dominates and we have $\sigma_L\approx constant$, as
expected. This second term is just the flat space contribution to the
quantum fluctuations, and we shall subtract it in what follows in order to
avoid the well known ultraviolet divergences in $<\Delta^2_{\lambda}>$.

Soon after a given wavelenth becomes larger than the horizon, it can be
treated as classical. Here we adopt the point of
view that Eq.(\ref{probability}) gives the probability distribution for the
initial amplitudes of the perturbations, which will evolve classically
thereafter.

{\em b- After inflation: while the loop is larger than the cosmological
horizon.} As we mentioned before, the loops are conformally stretched until
they enter the horizon. Since we do not have analytical solutions to
describe this epoch, we shall use the following approximations
\cite{vi81a,tuba84}: for wavelengths larger than the horizon the amplitude of
the perturbations is conformally stretched by the expansion,
whereas for wavelengths within the horizon the amplitude of the perturbations
remains constant.

Perturbations with wave number $L$ enter the horizon at time $t_L\sim
R(t_L)/L.$ At this time we have $\sigma_L=[2\mu L(L^2-1)]^{-1/2}HR(t_L)$.
The loop enters the horizon at the time
$$
t_c\sim R(t_c)\sim {R^2(t_L)\over t_L}\sim LR(t_L).
$$
At this time we still have the same $\sigma_L$ than at $t_L$, since
the amplitude of the perturbations is frozen after $t_l$, so we write
\beq
\sigma_L(t_c)=[2\mu L(L^2-1)]^{-1/2}HR(t_L)=[2\mu L^3(L^2-1)]^{-1/2}HR_c.
\label{jerusalem}
\eeq

{\em c- Loops inside the horizon.} Once a loop comes within the horizon we
can ignore the effects of the expansion of the Universe and consider the
collapse of a loop in flat space. In this case the evolution of the
perturbations can be solved analytically. The details can be found in
Appendix A. We find that during this period the behaviour of radial
perturbations $\Delta_r$ is different from that of transverse perturbations
$\Delta_t$. The amplitude of radial perturbations shrinks by a factor of
$L$ as the loop collapses, while the amplitude of transverse perturbations
remains constant. For $R<<R_c$ we find
\beq
\sigma_{r,L}\approx(2\mu L^7)^{-1/2}HR_c,\label{yoga}
\eeq
$$
\sigma_{t,L}\approx(2\mu L^5)^{-1/2}HR_c.
$$
Introducing (\ref{yoga}) in (\ref{probability}), we obtain the
probability distributions for the variables $\Delta^{(i)}_{\lambda,L}$ at
the time when the radius of the unperturbed loop enters the Schwarzschild
radius $R_s$.

The dynamics of gravitational collapse and
black hole formation from a given loop configuration is, of
course, a complicated issue that cannot be treated in detail analytically.
However, we know that if the amplitude of the perturbations is much smaller
than $R_s$ at the time $t_s$ when the unperturbed loop enters the
Schwarzschild radius, then a black hole will form. On the contrary, if the
perturbations are large compared with $R_s$ the loop will probably self
intersect and fragment into smaller loops
before it shrinks to a size $R_s$ (the resulting loops will be far from
circular, so a black hole will never form in this case). To compute the
probability of black hole formation, we shall count a particular loop as a
black hole with gaussian weight
\beq
e^{-\delta^2/\alpha^2 R_s^2}.\label{window}
\eeq
Here $\delta$ is the r.m.s. perturbation when one averages over the
circunference of the loop (at time $t_s$)
$$
\delta^2\equiv{1\over 2\pi}\int_0^{2\pi}\Delta^2(\theta,t_s)=
{1\over 2\pi}\sum_{i,\lambda,L}(\Delta^{(i)}_{\lambda,L})^2,
$$
where we have used (\ref{fourier}). In Eq. (\ref{window}), $\alpha$ is an
unknown parameter
of order 1, which models our lack of knowledge on the dynamical details.
We should emphasize that the precise form of the ``Gaussian window''
(\ref{window}) is not very important here; we have taken the Gaussian shape
just for computational convenience. Had we taken, for instance, a step
function window, our final conclusions would be the same.

{}From (\ref{probability}) and (\ref{window}), the probability of black
hole formation from a loop that has nucleated during inflation is given by
\beq
{\cal P}_{bh}=\int |\Psi|^2e^{-\delta^2/(\alpha R_s)^2}\prod_{\lambda,i,L}
d\Delta^{(i)}_{\lambda,L}=
\prod_{\lambda=r,t}\prod_{L=2}^{\infty}
\left(1+{\sigma_{\lambda,L}^2\over\pi\alpha^2 R_s^2}\right)^{-1}.
\label{mico}
\eeq
Using (\ref{yoga}), we can rewrite it as
\beq
{\cal P}_{bh}(L_*)=\prod_{L=2}^{\infty}
\left(1+2{L_*^5\over L^5}\right)^{-1}
\left(1+2{L_*^5\over L^7}\right)^{-1},\label{probh}
\eeq
where we have introduced the notation
$$L_*\equiv[16\pi^2B(\alpha G\mu)^2]^{-1/5}.$$
Here $B\equiv 4\pi\mu/H^2$ is the Euclidean action of the instanton
describing the nucleation of the loop. Note that ${\cal P}_{bh}$ does not
depend on $R_c$, the radius of the loop at horizon crossing. Note, also,
that the dependence on the parameters $H,\mu$ and $\alpha$ is only through
the combination $L_*$.

The physical meaning of $L_*$ is the following. Modes with $L>>L_*$ have
$\sigma_L<<\alpha R_s$. These modes will always have amplitudes much
smaller than the Schwarzschild radius and they will not contribute to
${\cal P}_{bh}$. This is also clear from (\ref{probh}). Therefore $L_*$ is
basically the number of modes whose amplitude can be large enough to
prevent the formation of a black hole.

The function ${\cal P}_{bh}(L_*)$ cannot be given in closed form. However,
one can compute it numerically to arbitrary precision by including a
sufficient number of terms in the product (\ref{probh}). A plot of
${\cal P}_{bh}$ versus $L_*$ is given in Fig. 1. From the graph we see
that the probability decays (essentially) as an exponential function of
$L_*$.

In deriving Eq.(\ref{probh})
we have neglected damping processes, such as gravitational
radiation and friction. In principle these could  reduce the amplitude
of the perturbations and therefore increase the probability of
black hole formation. In
Appendix B we briefly discuss the damping mechanisms and the limits in which
they can be neglected. We find that damping can be safely ignored from the
calculation of ${\cal P}_{bh}$ provided that $G\mu$ is in the range
\beq
(\alpha^2 B)^{-2/9}\left({m_p\over M}\right)^{5/9}\lesim G\mu< 5\cdot
10^{-2}(\alpha^2 B)^{1/3},\label{range}
\eeq
where $m_p$ is the Planck mass and $M$ is the mass of the black hole. The
upper limit is due to gravitational radiation, which is important for heavy
strings, while the lower limit corresponds to friction due to the
surrounding matter, which is important for light strings.
In the next Section we will be interested in ${\cal P}_{bh}$ for black
holes of mass $M\sim 10^{19} m_p$, so Eq.(\ref{range}) leaves us with
a wide range of values of $G\mu$ for which (\ref{probh}) is valid.

\section{Observational constraints on loop nucleation}

During inflation, loops of size $\sim H^{-1}$ are produced at a constant
rate per unit physical volume. After nucleation these loops grow
like the scale factor, so
we expect that the number of loops with size $\sim R$ contained in a
volume $\sim R^3$ will be independent of $R$. That is, we expect a scale
invariant distribution of loops . The number
density distribution of loops will be given by \cite{baal91}
\beq
{dN\over dV}={\nu}{dR\over R^4}.
\label{scaleinvariant}
\eeq
(Actually, the distribution (\ref{scaleinvariant}) has a lower cut-off at
$R\sim H^{-1}$, since we do not have loops smaller than that. It also has an
upper
cut-off at $R\sim EH^{-1}$, where $E$ is the total e-folding factor since
the onset of inflation. See Ref.\cite{baal91} for details.)

The coefficient $\nu$ is the number of loops produced during an expansion
time $H^{-1}$ in a volume $H^{-3}$. This has been estimated in the
semiclassical approximation in Ref.\cite{baal91}, using the instanton methods
 \cite{co85}
\beq
\nu=Ae^{-B}.\label{semiclass}
\eeq
Here, as in Section \ref{formation}, $B=4\pi\mu/H^2$ is the euclidean action
of the instanton describing the nucleation of the string (which is just a
spherical worldsheet of radius $H^{-1}$). The coefficient $A$ has not been
calculated and we shall leave it as a free parameter here, presumably of order
one. Let us now consider various observational constraints that one can place
on the parameters of the model.

{\em a- Constraint from primordial black hole abundance.}

After inflation, the distribution of loops with $R>>t$ is still given by
(\ref{scaleinvariant}), since these loops continue to be stretched by the
expansion. Loops with size smaller than the cosmological
horizon, $R<<t$, are not stretched and are simply diluted by the expansion,
\beq
{dN\over dV}=\nu\left[{a(R)\over a(t)}\right]^3{dR\over R^4},\label{joquese}
\eeq
where $a$ is the cosmological scale factor.
For loops that enter the horizon during the radiation dominated era, we
have
\beq
{dN\over dV}=\nu\left({1\over tR}\right)^{3/2}{dR\over R}.\label{much}
\eeq
Multiplying by the mass of the loops, $M\approx 2\pi\mu R$,
we obtain the mass density distribution
\beq
d\rho(M)=\nu\left({2\pi G\mu\over tGM}\right)^{3/2}dM.
\eeq
Dividing by the critical density $\rho_c=3/32\pi Gt^2$ we have, at the
present time,
\beq
d\Omega_s(M)\equiv {d\rho(M)\over\rho_c}\approx
{32\pi^2\sqrt{2\pi}\over 3}(G\mu)^{3/2}\nu
\left({t_{eq}\over GM}\right)^{1/2}{dM\over M},\label{domega}
\eeq
where $t_{eq}\approx 4\cdot 10^{10} sec.$ is the time of equal matter and
radiation densities.

As we described in Section \ref{formation}, some of the loops
in the network will form black
holes with probability given by eq.(\ref{probh}). To obtain the spectrum of
primordial black holes that are produced through this mechanism we
multipy (\ref{domega}) by ${\cal P}_{bh}$. The fraction of the density
parameter $\Omega$ in black holes of mass $\sim M$ is then given by
\beq
\Omega_{bh}(M)\approx {\cal P}_{bh}{32\pi^2\sqrt{2\pi}\over 3}
(G\mu)^{3/2}\nu\left({t_{eq}\over GM}\right)^{1/2}.\label{omega}
\eeq

The strongest observational constraint on the abundance of primordial black
holes comes from the emission of $\gamma$-rays by holes that are
evaporating at the present time \cite{maca91}. These black holes have a
mass $M\approx 5\cdot 10^{14} g$, and the constraint is given by
\beq
\Omega_{bh}(5\cdot 10^{14} g){\buildrel <\over\sim} 10^{-8}\label{obscon}.
\eeq
{}From (\ref{omega}) we have
\beq
{\cal P}_{bh}\cdot(G\mu)^{3/2}\nu{\buildrel <\over\sim} 10^{-28}.
\label{conundrum}
\eeq

We can rewrite this inequality as
\beq
-\log_{10}{\cal P}_{bh}(L_*)
-{3\over 2}\log_{10}(\alpha G\mu)+(\log_{10}e)B
{\buildrel >\over\sim }N,\label{conundrums}
\eeq
where $N=28+\log_{10}(A\alpha^{-3/2})$. Note that the left hand side
only depends on $B$ and the combination $\alpha G\mu$.
Therefore, for
given $N$, eq.(\ref{conundrums}) will exclude a certain region in the
plane $(B,\alpha G\mu)$.
Since the parameters $\alpha$ and $A$ are expected to be of order 1, their
contribution to $N$ will be small. Note also that even if the observational
constraint (\ref{obscon}) was improved by one order of magnitude, this
would only increase $N$ by one. At any rate, we expect $N\approx 28$ plus
or minus a few units.

In Fig. 2, the boundary of the excluded region in
parameter space is depicted for $N=27, 28$ and $30$. It is seen that the
result is not very sensitive to the value of $N$ and, quite generically,
the bound will be satisfied provided that
\beq
\alpha G\mu{\buildrel<\over\sim}10^{-4},\label{1}
\eeq
independently of $B$. This can be easily understood since,
for such values of $\alpha G\mu$,
the parameter $L_*$ will be very large and,
correspondingly, the probability of black hole formation will be
exponentially small (see Fig. 1).

Also, from Fig. 2 we see that
for large values of $B$, say $B{\buildrel>\over\sim} 50$, the constraint on
$G\mu$ is practically removed due to the exponential suppression in the
number of nucleated loops. Of course this limit is not very
interesting cosmologically.

{\em b-Constraints from the microwave background}

A string moving across the sky produces a discontinuity in the observed
temperature of the Cosmic Microwave Background (CMB) between both sides of the
string \cite{stebbins88}, roughly of order $(\delta T/T)\approx 10G\mu$.
For values of $\nu\sim 1$,
the loop sizes extend all the way to the present Hubble radius,
and the bound obtained in Ref. \cite{beal92}
$$
G\mu\lesim 10^{-6}, \quad (\nu\sim 1)
$$
should apply. However, for $\nu<<1$ this bound should not necessarily be
satisfied, since in this case all the loop radii are much smaller than
the horizon.

Let us first consider
the effect of loops with $R>t_{eq}$. The angle $\Theta$ subtended
by the string is given by
\beq
y\equiv\tan(\Theta/2)={R\over d_0}\left[1-{d_0\over 3 t_0}\right]^{-2},
\label{subtend}
\eeq
where $t_0$ is the age of the Universe and $d_0$ is the present distance to
the string. Loops of given radius start oscillating when $R\sim t$, and
therefore they will only
have an effect on the CMB provided that they are at a sufficiently low
redshift. Using the fact that the light that is now reaching us from a
point at a distance $d_0$ was emitted at time $t=t_0[1-(d_0/3t_0)]^3$,
loops that were oscillating when the observed
light passed through them must satisfy
\beq
R<t_0\left[1-{d_0\over 3 t_0}\right]^3.\label{murdstone}
\eeq

The number of loops with radius in the interval $[R, R+\Delta R]$ at a distance
in the interval $[d_0,\Delta d_0]$ is given by
\beq
\Delta N(R,d_0)={\nu\over R^2 t_0^2}4\pi d_0^2\Delta R\Delta d_0.
\eeq
Changing variables $\{R,d_0\}\to\{y,d_0\}$ through eq. (\ref{subtend}) and
integrating over all possible distances $d_0$ allowed by the constraint
(\ref{murdstone}) we obtain
\beq
dN(y)=36\pi\nu\left[{1\over 3y^3}-{1\over y^2}\ln\left(1+{1\over
3y}\right)\right]dy. \label{peggotty}
\eeq
The expectation value of the number of loops subtending an angle larger than
$\Theta=2\tan^{-1}y$ is then
$$
N_{>\Theta}=\int^{\infty}_y dN[y]=36\pi\nu\left({1\over 6y^2}+{1\over y}-
3\left[1+{1\over 3y}\right]\ln\left[1+{1\over 3y}\right]\right).
$$
This function is plotted in Fig. 3. It is seen that for $\Theta>10^{\circ}$
(the angular resolution of the COBE experiment) and $\nu<10^{-3}$,
for instance, the expected number of events $N_{>\Theta}$ is only of order 1.
This means that for such low values of $\nu$ we can hardly expect to see
any loops that large.

However, one must take into account that `unresolved' loops whose
angular size is smaller than the detector beam angular size
can still produce a signal, although of lower magnitude. For
simplicity, in what follows we shall assume that the loops are nearly
circular. In that case, if the plane of the loop is perpendicular to the
line of sight, the temperature distortion in the CMB has the profile of a
`top hat' with radius equal to the radius of the loop at the time when the
light rays crossed this plane \cite{stebbins88}. On the angular scale
set by $R$ (the radius of the loop when it is at rest),
the temperature fluctuation is $|\delta T/T|\approx 10G\mu$ (the
sign depending on whether the loop is expanding or contracting at the time
when the light rays cross the loop). On larger angular scales, the signal
will be inversely proportional to the solid angle that we are considering.

Therefore
if the detector beam has an angular size $\beta$ larger than the angular
size of the loop $\Theta(R,d_0)$, the measured temperature fluctuation
will be
\beq
\Delta\equiv{1\over 10G\mu}{\delta T\over
T}\approx\left(\Theta\over\beta\right)^2.\label{emly}
\eeq
For small angles, the distribution (\ref{peggotty}) can be approximated by
$dN(\Theta)=48\pi\nu \Theta^{-3} d\Theta$, and using (\ref{emly}) to
express $\Theta$ as a function of $\Delta$ we have
$$
N_{>\Delta}={24\pi\nu\over \beta^2\Delta}.
$$
Here $N_{>\Delta}$ is the expected number of events due to unresolved loops
that give a signal larger than $\delta T/T$ on a detector with beam angular
size
$\beta$. This equation can be rewritten as
\beq
\nu G\mu\approx {N_{\Delta}\over 240\pi}\beta^2{\delta T\over T},\label{ham}
\eeq
where $\beta$ has to be expressed in radians.

The COBE experiment constrains $\delta T/T$ to be less than $10^{-5}$
on scales $\beta\approx 10^{\circ}\approx .17\ rad$. It is not clear what
is the maximum number of events $N_{\Delta}$ allowed by the COBE data.
Taking $N_{\Delta}\lesim 1$ seems a little over restrictive since one
small string can easily be hidden behind the galaxy. To be more
conservative we shall take $N_{\Delta}\lesim 10$, which from (\ref{ham}),
yields the bound
\beq
\nu G\mu\lesim 10^{-8}. \quad    (\nu<10^{-2}) \label{2}
\eeq
This bound comes from the contribution of unresolved
loops, and it
only applies for $\nu<10^{-2}$. For larger values of $\nu$, the number of
`resolved' loops is sufficiently large that the usual bound
\beq
G\mu\lesim 10^{-6}    \quad      (\nu>10^{-2}) \label{3}
\eeq
should be used.

So far we have considered loops with $R>t_{eq}$. A similar analysis can be
carried out for $R<t_{eq}$ to show that these loops cannot produce
relative temperature fluctuations in excess of $10^{-5}$ provided that
the constraints (\ref{2}), (\ref{3}), and Eq.(\ref{1}) from the previous
subsection are satisfied.

{\em c-Constraint from gravitational radiation}

The millisecond pulsar observations place a constraint on the density
parameter in gravitational radiation of period $\sim$ 1 year
\cite{stal90},
\beq
\Omega_g<4\cdot 10^{-7} h^{-2}.\label{milisec}
\eeq
{}From strings, we have \cite{vavi85}
$$
\Omega_g\sim {128\pi\over9}\nu\left({G\mu\over\gamma_g}\right)^{1/2}\Omega_r
$$
where $\Omega_r\sim4\cdot10^{-5}h^{-2}$ is the present density parameter in
radiation and $\gamma_g\sim 100$. With this, the bound (\ref{milisec}) reads
$$
\nu(G\mu)^{1/2}\lesim 2\cdot10^{-3}.
$$
This is always satisfied provided that the bounds (\ref{1}),(\ref{2}) and
(\ref{3}) from the previous subsections are satisfied.

{\em c-Structure formation}

A question of cosmological interest is whether the constraints derived so
far are compatible with a scenario in which the nucleated strings would
seed the formation of large-scale structure in the Universe. Let us make a
rough
estimate of the values of the parameters needed for such purpose.

Assuming, for simplicity, cold dark matter, the mass accreted by a loop of
radius $R<t_{eq}$ at the present time is
\beq
M=2\pi\mu R z_{eq}.\quad (R<t_eq)\label{massa1}
\eeq
Here $z_{eq}\approx 2\cdot 10^{-4} h^2$ is the redshift at time $t_{eq}$,
with $h$ the Hubble constant in units of $100km\ s^{-1}Mpc^{-1}$.
This is because perturbations start growing at time $t_{eq}$ and they grow
proportionally to the scale factor.
{}From (\ref{much}) the number density of loops of radius $\sim R$ is
$n_R\sim \nu(t_{eq} R)^{-3/2}z_{eq}^{-3}$. Using (\ref{massa1}), the mean
separation between objects of mass $M$ is given by
\beq
d_M\sim n^{-1/3}_{R(M)}\sim \nu^{-1/3}\left({Mt_{eq}z_{eq}\over
2\pi\mu}\right)^{1/2}.\label{yougotit}
\eeq

Rich clusters of mass $M_{cl}\sim 10^{15}h^{-1}
M_{\odot}$ are separated by distances of order $d_{cl}\sim 50 h^{-1} Mpc$.
Therefore,
from (\ref{yougotit}), we obtain the normalization
\beq
\nu^{2/3}G\mu\approx 5\cdot10^{-8}h^{-1}.\quad(\nu\lesim
10^{-2})\label{normal1}
\eeq
This normalization is valid for $G\mu\gesim 10^{-6} h$
(hence $\nu\lesim 10^{-2}$). For smaller values of $G\mu$ the loops that
accrete masses of order $M_{cl}$ have $R>t_{eq}$, and (\ref{massa1}) does
not apply.

For $G\mu\lesim10^{-6} h$, loops start accreting matter when they enter the
horizon, and we have
$$
M=2\pi\mu\nu R\left({t_0\over R}\right)^{2/3}.\quad (R>t_{eq}).
$$
Using $n_R=\nu/(Rt_0^2)$ we obtain
$$
d_M\sim n^{-1/3}_{R(M)}\sim\nu^{-1/3}M(2\pi\mu)^{-1}.
$$
Again, normalizing for rich clusters, we find
\beq
\nu^{1/3}G\mu\approx 2\cdot 10^{-7}.\quad (\nu\gesim 10^{-2})\label{normal2}
\eeq

The region excluded by the constraints (\ref{1}), (\ref{2}) and (\ref{3})
in parameter space is plotted
schematically in Fig. 4 (shaded region). It is seen that
the parameters needed for structure formation (thick solid line)
[from (\ref{normal1}) and (\ref{normal2})] lie marginally in the allowed
region.
Of course, the normalizations (\ref{normal1}) and (\ref{normal2}) should
not be taken too literally, and one expects large error bars in the
thick solid line of Fig. 4. However, a scenario in
which  the nucleated strings may seed some of the observed structure in our
universe does not seem to be ruled out.

We should now mention the effects of compensation \cite{vest90} which has
been ignored in the above discussion. When a loop nucleates, its energy is
balanced by a corresponding deficit in the local densities of matter and
radiation. This compensation reduces the gravitational effect of the loop
on surrounding matter and on background photons on scales greater than $R$.
When a compensated loop radiates away its energy, it leaves behind an
underdense region. As the density contrast grows, this region will evolve
into a void with a dense central object seeded by the loop. The
compensation is somewhat reduced by radiation and
neutrinos streaming into the underdense regions and can be further reduced
by loop fragmentation. The effect of compensation on structure formation
has not been fully investigated, but it seems reasonable to assume that
this effect will not be dramatic on scales crossing the horizon before
$t_{eq}$. Then the relations (\ref{normal1}),(\ref{normal2}) should still
be valid by order of magnitude.

\section{Domain walls and monopoles}

Just as in the case of strings, spherical domain walls can spontaneously
nucleate at a constant rate per unit volume during inflation. As a result,
the Universe will get filled with a scale invariant distribution of walls
given by (\ref{scaleinvariant}) and (\ref{semiclass}), where now
$B=2\pi^2\sigma H^{-3}$ is the action of the instanton describing the
nucleation of the wall.

The observational constraints on domain wall nucleation are very different
from the ones that one can impose on strings, since the mass of a spherical
wall
\beq
M=4\pi\sigma R^2 \label{tuquesaps}
\eeq
grows quadratically rather than linearly with the
radius. An inmediate consequence is that  walls of size
\beq
R_c>(8\pi G\sigma)^{-1}
\label{alucina}
\eeq
will all collapse to form black holes \cite{baal91}, since for them
$R_s>R_c$.

To estimate the microwave background anisotropy induced by domain wall
bubbles, one has to take propper account of the compensation effect
discussed at the end of the previous section. This is a somewhat complicated
issue and we shall
not attempt to address it in detail here. However, one can make a rough
estimate of what this effect should be by using the following arguments

In a pure dust universe the
compensation would be complete, and background photons would be
unperturbed on scales greater than the co-moving scale of $R_s$ at horizon
crossing. Therefore, on larger scales, the whole effect should be due to
the underdensity in radiation that is needed to compensate for the mass of
the black hole.
This underdensity causes radiation from neighboring regions to move in,
partially destroying compensation. As this happens, the initial
underdensity propagates away from the black hole as a sound wave at the
speed of light.
At time $t_{eq}$ the mass of the underdensity, $M_r$, is
comparable to the mass of the black hole, $M_{bh}$. At later times, we have
$M_r\approx M_{bh}z/z_{eq}$, due to cosmological redshift. The
gravitational potential of $M_r$ extends up to scales of order $t$, and we
have $\phi\sim GM_rt^{-1}$. This will cause temperature distortions of
order
$$
{\delta T\over T}\sim\phi\sim {G M_{bh}\over t_0}{z^{5/2}\over z_{eq}},
$$
on scales of the order of the cosmological horizon at redshift $z$.

The angle $\alpha$ subtended by the horizon at redshift $z$ is given by
$\tan\alpha\approx(z^{1/2}-1)^{-1}$, so at small angles $\delta T/
T\propto\alpha^{-5}$. In particular, for $\alpha\approx 10^{\circ}$, we
have
\beq
{\delta T\over T}\sim {M_{bh}\over M_u}\quad (\alpha\approx 10^{\circ}),
\label{yarmouth}
\eeq
whereas for $\alpha\approx 90^{\circ}$ we have a much lower effect
$\delta T/ T\sim M_{bh}/(M_uz_{eq})$. Here $M_u\equiv G^{-1}t_0$ is
approximately the mass of our observable universe.

Let us denote by $R_{max}$ the radius of the largest wall that may have
existed in our observable universe (up to distances $d_0=3 t_0$). How
massive can this wall be? From (\ref{yarmouth}), and using $\delta T/
T\lesim 10^{-5}$, we have
\beq
M_{max}\lesim 10^{-5}M_u\approx 10^{52} g\label{quinion}.
\eeq

{}From (\ref{tuquesaps}) and (\ref{quinion})
we can see that $R_{max}$ must be less than $t_{eq}$
even if $\sigma$ is as low as the electroweak scale.
Using
$$
{dN\over dV}={\nu\over R^{5/2}t_{eq}^{3/2}}\left({t_{eq}\over t_0}\right)^2
dR,
$$
the number of walls with size larger than $R$ within a volume
$4\pi/3(3t_0)^3$ is
$$
N_{>R}=24\pi\nu{t_{eq}^{1/2}t_0\over R^{3/2}}.
$$
Therefore, the larger wall that we can expect to find in the
observable universe has radius
$R_{max}\approx [24\pi\nu t_eq^{1/2}t_0]^{2/3},$ and  Eq. (\ref{quinion})
yields the bound
\beq
\nu^{4/3}{\sigma\over m_p^3}{\buildrel<\over\sim} 10^{-64}.\label{fot}
\eeq

Let us now turn to monopole pair production during inflation.
Unlike walls and strings, monopoles are not stretched to enormous sizes by
the inflationary expansion. They
are just diluted. An inmediate consequence is that the only monopoles
that will be relevant at the end of inflation are the ones that have been
produced during the last expansion time. This density is given by
\cite{baal91}
\beq
n=H^3 A e^{-{2\pi m/H}},\label{ous}
\eeq
where $m$ is the mass of the monopole (here, as in the previous sections,
the exponent $2\pi m H^{-1}$ is the action of the relevant instanton.)

A constraint on the abundance of nucleated monopoles comes from
the fact that they should not recreate the monopole problem that inflation
was aimed to solve. Of course this can be easily achieved by choosing the
ratio $m/H$ to be sufficiently large, but as we shall see, one does not
necessarily have to impose that.

Actually, the number density of monopoles at the time
of reheating depends very much on the details of how inflation ended.
Consider, as an example, a model in which the usual exponential expansion
is followed by a short period of power law expansion $a\propto t^p$ that
starts at time $t_1$ and ends at time $t_2$ ($p$ may be larger that 1, in which
case we have power law inflation, but it need not be).
At time $t_2$ reheating is
completed and we enter the usual radiation dominated era.

It is clear that, if no monopoles are created during the power law epoch,
we will have, at time $t_2$,
\beq
n(t_2)=n(t_1)\cdot f^{3p},\label{lio}
\eeq
where $n(t_1)$ is given by (\ref{ous}) and
$$
f\equiv\left({t_1\over t_2}\right)\approx 1.6{\cal N}^{1/2}
{T_r^2\over H m_p}.
$$
Here ${\cal N}$ is the effective number of massless degrees of freedom at
reheating and we have used $t_1=pH^{-1}$ and $t_2=2p\ m_p(10{\cal N})^{-1/2}
T_r^{-2}$, with $m_p$ the Planck mass and $T_r$ the reheating temperature.
Even if we assume that monopoles
continue to be produced at a rate per unit volume
given by
$$
A {\cal H}^4 e^{-{2\pi m\over {\cal H}}},
$$
where ${\cal H}\equiv \dot a/a$ is the instantaneous expansion rate,
it is not
difficult to show (for $2\pi m H^{-1}>3p-4$) that the density of
monopoles during the power law era is dominated by the ones that were
already present at $t_1$. Therefore Eq.(\ref{lio}) is still valid in this case.

After reheating, the monopoles continue to be diluted as $a^{-3}$ whereas
the density in radiation decays as $a^{-4}$. Therefore
the contribution of the
monopoles to the density parameter at the time $t_{eq}\approx 10^{10} s$
given by
$$
\Omega_m={m\cdot n(t_2)\over \rho_c(t_2)}\left({t_{eq}\over t_2}\right)^{1/2}
\approx 10^{29}f^{3[p-{1\over 2}]}\left[{H\over m_p}\right]^{3/2}
A e^{-{2\pi m\over H}} {m\over m_p}.\label{traddles}
$$

The Parker bound \cite{pa79}
requires $\Omega_m{\buildrel<\over\sim} 10^{-4} (m/m_p)$
(assuming Dirac charges for the monopoles), so we must impose
$$
\left[{H\over m_p}\right]^{3/2}f^{3[p-{1\over 2}]}Ae^{-{2\pi m\over H}}
{\buildrel <\over\sim} 10^{-33}.
$$
As mentioned above,
this bound can be trivially satisfied by supressing the creation
of monopoles with a sufficiently large $m/H$ ratio. Alternatively, if
$p>1/2$, then $\Omega_m$ will decrease with the reheating temperature
through the supression factor $f$. For instance, taking $p=3.5$ and
$f\sim 10^{-4}$ the bound is automatically satisfied without any assumptions
on $m/H$.

\section{Conclusions}

In this paper we calculated the probability of black hole formation by a
string loop spontaneously nucleated during inflation. The calculation was
based on the quantum theory of perturbations on strings in de Sitter space
developed in Ref.\cite{gavi92}. The result is given in Eq.(\ref{probh}) and
in Fig. 1. We then used the probability (\ref{probh}) to derive
observational constraints on the mass parameter of the string and on the
loop density parameter $\nu$. The strongest constraint comes from the
emission of $\gamma$-rays by evaporating black holes. Additional
constraints are due to the absence of characteristic hot and cold spots
that would be produced by large oscillating loops on the microwave sky. The
gravitational radiation background generated by the loops imposes no new
constraints. The excluded region of the parameter space is sketched in Fig.
4.

We have also briefly discussed the cosmological implications of nucleated
domain wall bubbles and monopole-antimonopole pairs. For sufficiently large
bubbles, the formation of black holes is inevitable. Although these black
holes can have enormous masses, their effect on structure formation and on
the background radiation anisotropy is diminished due to the compensation
effect \cite{vest90}. The predicted density of the monopoles is
given by Eq.(\ref{traddles}). Depending on the values of the parameters, it
can be too high, negligible, or it can lie in the interesting range and be
potentially observable.

\section*{Acknowledgements}

This research was supported in part by the National Science Foundation
under grants N$^\circ$ PHY 89-04035, PHY 87-14654 and
J.G. is supported by a Fulbright grant.
A.V. is grateful to Sidney Coleman and to the High Energy Theory group at
Harvard for their hospitality.

\appendix
\section*{Appendix A}
\setcounter{equation}{0}
\renewcommand{\theequation}{A\arabic{equation}}

In this Appendix we study linear perturbations to a circular loop
of string in flat space. Before going to the particular case of a circular
loop, we shall consider
perturbations to an arbitrary string configuration whose worldsheet is
given by $x^{\mu}(\xi^a)$ ($\xi^a$ are arbitrary coordinates on the
worldsheet). This will be a trivial extension of the general
results given in Refs. \cite{gavi91,gavi92} for the case of domain walls.

Following Ref. \cite{gavi91}, we parametrize the perturbed
worldsheet $\tilde x^{\mu}$ as
\beq
\tilde
x^{\mu}(\xi)=x^{\mu}(\xi)+\sum^2_{A=1}n^{A\mu}\phi^A,\label{perturbed}
\eeq
where $n^{A\mu}$ are the two vectors normal to the unperturbed worldsheet.
They satisfy
\beq
n^{A\mu}\partial_ax_{\mu}=0, \quad n^{A\mu}n^B_{\mu}=\delta^{AB}.
\label{satisfy}
\eeq
The perturbation fields $\phi^A$ have the meaning of normal displacements
to the worldsheet, as measured by an observer that is moving with the
unperturbed string.

The dynamics of the string is governed by the Nambu action
\beq
S[x^{\mu}(\xi)]=-\mu\int\sqrt{-g}\ d^2\xi,\label{nambu}
\eeq
where $g$ is the determinant of the  metric induced on the worldsheet
$g_{ab}=\partial_{a}x^{\mu}\partial_bx_{\mu}$. The equations of motion that
result from (\ref{nambu}) are well known
\beq
\Box x^{\mu}(\xi^a)=0,\label{wellknown}
\eeq
where $\Box$ stands for the covariant d'Alembertian on the worldsheet.
Multiplying (\ref{wellknown}) by $n^A_{\mu}$ and integrating by parts one
obtains
\beq
g^{ab}K_{ab}^A=0,\label{newform}
\eeq
where $K_{ab}^A\equiv-\partial_an^A_{\mu}\partial_bx^{\mu}$ is the
extrinsic curvature corresponding to the normal $n^A$.

The effective action for the perturbation fields $\phi^A$ on a given
background $x^{\mu}(\xi)$ can be obtained by introducing (\ref{perturbed}) into
the action (\ref{nambu}) and then expanding to quadratic order in $\phi^a$.
After some lengthy algebra, the result can be written as
$$
S[\tilde x^{\mu}]=S[x^{\mu}]+S_{\phi}.
$$
The first term is just the action for the unperturbed solution, while the
second is given by
\beq
S_{\phi}=-\mu\int\sqrt{-g}\left[{1\over 2}\phi^A_{,a}\phi^A,^a-
{1\over 2}K_{ab}^AK^{Bab}\phi^A\phi^B+{\cal S}\right] d^2\xi,\label{sphi}
\eeq
with
$$
{\cal S}={1\over 2}\phi^A\phi^B n^{C\mu}n^{C\nu}\partial_an^A_{\mu}
\partial^an^B_{\nu}+\phi^A_{,a}\phi^Bn^{A\mu}n^B_{\mu},^a.
$$
In deriving this effective action we have used the equations of motion
(\ref{newform}) to eliminate the terms linear in $\phi^A$. From
(\ref{sphi}) we see that, in general, the two fields $\phi^A$ are coupled
to each other in a complicated way.

Particularizing to the case of a circular loop, matters will simplify
considerably. The unperturbed worldsheet in cartesian coordinates
$(t,x,y,z)$ is given by
\beq
x^{\mu}=(t,R_c \cos\theta\cos{t\over R_c},R_c \sin\theta\cos{t\over
R_c},0).\label{circularloop}
\eeq
Here $R_c$ is just the radius of the loop at $t=0$ (when the loop is at
rest). The two vectors normal to the worldsheet, can be chosen as
$$
n_{\mu}^{(1)}={1\over\cos{t\over R_c}}(\sin{t\over
R_c},\cos\theta,\sin\theta,0),
$$
\beq
n_{\mu}^{(2)}=(0,0,0,1).\label{normals}
\eeq
The first one is a radial normal vector, while the second one is transverse
to the plane of the loop.
It is easy to see that in this case $K_{ab}^{(2)}=0$ and ${\cal S}=0$, so
the fields $\phi^{(1)}$ and $\phi^{(2)}$ decouple from each other. They
simply behave like free scalar fields in the curved geometry of the
unperturbed worldsheet.

In particular, the transverse perturbation behaves like a massless
minimally coupled field. The corresponding equation of motion
\beq
\Box
\phi^{(2)}={1\over\sqrt{-g}}\partial_a(\sqrt{-g}g^{ab}\partial_b
\phi^{(2)})=0\label{trivially}
\eeq
can be solved trivially. The metric on the worldsheet is given by
$$
ds^2_{\Sigma}=g_{ab}d\xi^a d\xi^b=\cos^2(t/R_c)
[-dt^2+R^2_cd\theta^2].
$$
Introducing this metric in (\ref{trivially}) we obtain a flat-space wave
equation
\beq
\ddot\phi^{(2)}-{1\over R^2_c}\phi''^{(2)}=0,\label{ordinary}
\eeq
where a prime denotes derivative with respect to $\theta$.
Eq.(\ref{ordinary}) is not surprising, since in 1+1 dimensions the
conformal coupling is the same as the minimal coupling
(see e.g. \cite{bida82}).
The mode solutions of (\ref{ordinary}) are just standing-waves that
oscillate with constant amplitude. Note that $\phi^{(2)}$ coincides with
$\Delta_r$ of Section \ref{formation}, so we have the following result:
{\em transverse perturbations to a circular loop oscillate with constant
amplitude as the loop collapses.}

The equation of motion for the radial perturbations $\phi^{(1)}$ can
be found from (\ref{sphi}) (with $K_{ab}^{(2)}={\cal S}=0$). We have
\beq
\Box\phi^{(1)}+K^{(1)}_{ab}K^{(1)ab}\phi^{(1)}=0.\label{nasty}
\eeq
The extrinsic curvature can be obtained from (\ref{circularloop}) and
(\ref{normals}):
$$
K_{tt}^{(1)}={1\over R_0},\quad K_{\theta\theta}^{(1)}=R_0,\quad
K_{t\theta}^{(1)}=0.
$$

The proper perturbation $\phi^{(1)}$, as measured by a local observer that
is moving with the string, is related to the radial perturbation $\Delta_r$
defined in Section \ref{formation} through a Lorentz contraction factor
$$
\Delta_r=\phi^{(1)}\sqrt{1-\dot R^2}.
$$
Writing (\ref{nasty}) in terms of $\Delta_r$ and decomposing in Fourier
modes [as in (\ref{fourier})] we have, after some algebra,
$$
\ddot\Delta_{r,L}+{2\over R_c}
\tan\left({t\over R_c}\right)\dot\Delta_{r,L} +{(L^2-1)\over
R_c^2}\Delta_{r,L}=0.
$$
The general solution to this equation can be written as
$$
\Delta_{L}(t)={\Delta_L(0)\over L}
\left[\sqrt{1-{R^2\over R_c^2}}\sin{Lt\over R_c}+L{R\over R_c}\cos{Lt\over
R_c}\right]+
$$
\beq
\dot\Delta_L(0){R_c\over 1-L^2}
\left[\sqrt{1-{R^2\over R_c^2}}\cos{Lt\over R_c}-L{R\over R_c}\sin{Lt\over
R_c}\right],\label{mongo}
\eeq
where $\Delta_L(0)$ and $\dot\Delta_L(0)$ are the values of the
perturbation and its derivative at $t=0$ (we have dropped the subindex
$r$), and $R=R_c\cos(t/R_c)$ is the radius of the unperturbed loop.

In the cosmological problem that we are interested in, $t=0$ corresponds to
the moment when the loop enters the horizon and starts collapsing. The
probability distribution for the initial conditions $\Delta_L(0)$ will be
given by (\ref{probability}) and (\ref{jerusalem}). We also need the
initial conditions $\dot\Delta_L(0)$. When the wavelength of a perturbation
is within the horizon we expect $<\dot\Delta^2_L>=(L/R_c)^2<\Delta_L^2>$,
where the brakets indicate average over one oscillation period. Therefore,
by the time the loop enters the horizon, the perturbations will have
developed velocities of order $\dot\Delta(t=0)\sim(L/R_c)\Delta_L(t=0)$.
Using this in (\ref{mongo}) we have
$$
\Delta_L(R<<R_c)\approx{\Delta_L(R_c)\over L}.
$$

That is to say, {\em the radial perturbations shrink by a factor of $L$ as
the loop collapses}.

\appendix
\section*{Appendix B}
\setcounter{equation}{0}
\renewcommand{\theequation}{B\arabic{equation}}

Damping mechanisms, such as gravitational radiation and friction, may
decrease the amplitude of the perturbations on a circular loop of string
and therefore increase the probability of black hole formation. In this
appendix we briefly discuss the limits in which the effects of gravitational
radiation and friction can be ignored.

Gravitational radiation can only smooth out perturbations on wavelengths
smaller than $\lambda_g\equiv\gamma_g G\mu t$, where $\gamma_g\sim 100$
(see e.g. \cite{vi85}). Taking $t\sim R_c$, this corresponds to wave
numbers larger than $L_g\equiv 2\pi(\gamma_g G\mu)^{-1}$. It is clear,
from the physical interpretation of $L_*$ discussed in Section
\ref{formation}, that
gravitational damping can be neglected in the calculation of ${\cal P}_{bh}$
providing that $L_g>L_*$. This inequality gives
\beq
G\mu<(2\pi\gamma_g^{-1})^{5/3}(16\pi^2\alpha^2 B)^{1/3}\approx
(17\gamma_g^{-1})^{5/3}(\alpha^2 B)^{1/3}.\label{pork}
\eeq
Typically, the right hand side of (\ref{pork})
will be of order $10^{-2}$ or larger, so
this is not a very strong condition.

Similarly we can consider the effects of friction. The dominant
contribution to friction comes from Aharonov-Bohm scattering of ambient
particles off the string in the radiation dominated era (see
Ref.\cite{vi91}). Assuming a situation in which the wavelength of the
perturbations is conformally stretched by the expansion of the universe
(as it is in the present case), it is shown in Ref.\cite{gasa92} that
friction can only be important for wavelengths smaller than
$\lambda_f\equiv \gamma_f(G\mu T)^{-1}$. Here T is the temperature and
$\gamma_f$ is a numerical coefficient of order one. The physical reason is
that for $\lambda>\lambda_f$, the friction term in the equations of motion
will ``switch off'' before the perturbations cross the cosmological horizon
and start to oscillate.

The wave number corresponding to $\lambda_f$ is
$$
L_f\equiv 2\pi\gamma^{-1}_f RTG\mu=\gamma^{-1}_fGMT_c,
$$
where $T_c$ is the temperature of the Universe at the time $t_c$ when the
loop crosses the horizon, $M$ is the mass of the loop at $t_c$ (which is
also the mass of the resulting black hole) and we have used that
the product $RT$ is independent of
time before the loop crosses the horizon. Using
$T_c\approx(10G{\cal N})^{-1/4}t_c^{-1/2}$ and
$t_c\approx R_c=M(2\pi\mu)^{-1}$(where ${\cal N}\sim 10^2$ is the effective
number of massless degrees of freedom) we have
$$
L_f\equiv (10{\cal N})^{-1/4}\gamma_f^{-1}(2\pi G\mu)^{1/2}
\left({M\over m_p}\right)^{1/2}.
$$
Similarly to the case of gravitational radiation, friction can be
neglected in the calculation of ${\cal P}_{bh}$ providing that $L_f>L_*$.
This condition is equivalent to
\beq
G\mu\gesim
\left({{\cal N}\gamma_f^4\over 230}\right)^{5/18}(\alpha^2 B)^{-2/9}
\left({m_p\over M}\right)^{5/9}.\label{truja}
\eeq
Therefore, the range of values of $G\mu$ for which friction can be ignored
depends on the mass of the black hole that one wishes to consider.

Putting (\ref{pork}) and (\ref{truja}) together and taking
$\gamma_f\sim 1$,$\gamma_g\sim 10^2$, we conclude that damping is unimportant
for values of $G\mu$ in the range
$$
(\alpha^2 B)^{-2/9}\left({m_p\over M}\right)^{5/9}\lesim G\mu< 5\cdot
10^{-2}(\alpha^2 B)^{1/3}.
$$

\section*{Figure captions}
\begin{itemize}
\item {\bf Fig. 1} The function ${\cal P}_{bh}(L_*)$ can be computed
numerically from Eq.(\ref{probh}) to arbitrary precision. We see that,
essentially, the probability of black hole formation decays as an
exponential function of $L_*$.

\item {\bf Fig. 2} We represent the boundary of the region excluded by
the constraint (\ref{conundrums}) in parameter space ($\alpha G\mu$ versus
$B$)
for $N=27$ (dashed line), $N=28$ (solid line) and $N=30$ (dotted line).
The result is not very sensitive to the value of $N$ and, quite
generically, the bound is satisfied provided that $\alpha G\mu
{\buildrel <\over\sim} 10^{-4}$.

\item{\bf Fig. 3} A plot of the number of loops $N_{>\Theta}$ that would be
seen in the sky at an angular size $\Theta$.

\item{\bf Fig. 4} Schematic plot of the region excluded in parameter space
by the observational constraints on loop nucleation. The horizontal
boundary at $G\mu\approx 10^{-6}$ is due to distortions in the CMB caused
by loops with angular size larger than 10$^{\circ}$. The slanted part of
the boundary is due to `unresolved' loops, with angular size less than
$10^{\circ}$. The horizontal boundary at $G\mu\approx 10^{-4}$ comes from
limits on the abundance of primordial black holes. The thick line is meant
to represent values of the parameters that roughly satisfy the
normalizations (\ref{normal1}) and (\ref{normal2}), and which may be
adequate
for structure formation.

\end{itemize}

\end{document}